\documentstyle[aps]{revtex}
\input amssym.def
\begin{document}
\twocolumn

\title{
Quantum Gravity Without Ghosts
}
\author{
Bryce DeWitt$^{+}$
and
C. Molina-Par\'{\i}s$^{++}$
}
\address{
$^{+}${Department of Physics, University of Texas, Austin, Texas
78712}\\
$^{++}${Theoretical Division T-6 and T-8, Los Alamos National Laboratory,
Los Alamos, New Mexico 87545}
}
\date{\today}

\maketitle

{\small {\bf Abstract:}
An outline is given of a recently discovered technique for building a
quantum effective action that is completely independent of
gauge-fixing choices and ghost determinants. One makes maximum use of
the geometry and fibre-bundle structure of the space of field
histories and introduces a set of nonlocal composite fields: the
geodesic normal fields based on Vilkovisky's connection on the space
of histories. The closed-time-path formalism of Schwinger, Bakshi,
Mahantappa {\it et al} can be adapted for these fields, and a set of
gauge-fixing-independent dynamical equations for their expectation
values (starting from given initial conditions) can be computed. An
obvious application for such equations is to the study of the
formation
and radiative decay of black holes, and to other back-reaction
problems.}

The back-reaction  problem in quantum gravity is one of the most
difficult in theoretical physics, firstly because one does not know
how best to pose it, and secondly because quantum gravity is not a
perturbatively renormalizable theory. One believes that the second
difficulty can be overcome by means of string theory. But string
theory is not yet a fully coherent discipline.

What one {\it can} do at present is treat standard quantum gravity as
an {\it effective} field theory. Although this theory is not
perturbatively renormalizable it will generate a unitary S-matrix if
one performs dimensional regularization, and minimally subtracts the
poles that arise in each perturbation order. No renormalization group
is
obtained, and each choice of auxiliary regularization mass yields a
different theory at high energies. But by setting this mass equal to
the Planck mass, one should obtain results that give a fair
description
of the physics even close to the Planck regime. Indeed, in view of the
known utility of the WKB approximation in ordinary quantum mechanics,
one may expect to get useful results already in $1$-loop order. But
{\it what} results?

Computing an S-matrix is of no help in tackling the backreaction
problem.  What is needed is the {\it quantum effective action}
$\Gamma$, which generates the full vertex functions of the theory and
which in principle describes all quantum effects, not merely
scattering processes. But the functional form of $\Gamma$ is needed
off shell.

It is not hard to construct a $\Gamma$ that is diffeomorphism
invariant even off shell, but there are infinitely many ways of doing
this, each corresponding to a different choice of covariant
gauge-fixing term and associated ghost determinant in the functional
integral that defines $\Gamma$. Each choice leads to a different set
of nonlocal dynamical equations for the ``mean value'', say, of the
metric tensor $g_{\mu \nu}$. Therefore this ``mean value'' does not
mean much and cannot be relied upon to give unambiguous insights.

One needs a quantum effective action that, like the classical action,
yields dynamical equations that are ghost and gauge-fixing
independent. The authors have recently shown how such an effective
action can be built \cite{1}. One makes maximum use of the geometry of
the space ${\rm Riem}({\rm M})$ of (pseudo) Riemannian $4$-metrics on
spacetime ${\rm M}$. ${\rm Riem}({\rm M})$ may be viewed as a
principal fibre bundle having for its typical fibre the diffeomorphsim
group ${\rm Diff}({\rm M})$. Real physics takes place in the base
space ${\rm Riem} ({\rm M})/{\rm Diff}({\rm M})$ of this bundle. ${\rm
Riem}({\rm M})$, viewed as an infinite dimensional manifold, may
itself be endowed with a diffeomorphism-invariant Riemannian
metric. If this metric is required to be {\it ultralocal} (a
requirement that is essential for the success of any renormalization
program) it is not difficult to show that it must belong to the
one-paramenter family
\begin{eqnarray}
\gamma^{\mu \nu \sigma ' \tau '}&=& {1 \over 2} g^{1/2}
(g^{\mu \sigma} g^{\nu \tau} + g^{\mu \tau} g^{\nu \sigma}+
\lambda g^{\mu \nu} g^{\sigma \tau})
\delta (x,x'), 
\nonumber\\
 g&=&|{\rm det}(g_{\mu \nu})|
\; .
\label{(1)}
\end{eqnarray}

A condensed notation is convenient here \cite{3}.  The metric field, or any
gauge field, will be denoted by ${\varphi}^i$, indices now being
understood as ranging over a continuum of values that specify
discrete components as well as spacetime points $x,x'
\dots$ . Summations over repeated indices then include integrations
over spacetime. Infinitesimal gauge transformations (diffeomorphisms)
may be expressed in the form
\begin{eqnarray}
\delta \varphi^i&=&Q^i_{\;\alpha} \delta \xi^\alpha
\; ,
\label{(2)}
\end{eqnarray}
where the $\delta \xi^\alpha$ are infinitesimal parameters and the
$Q^i_{\;\alpha}$ are components (in the ``coordinates'' $\varphi^i$)
of an infinite set of vector fields ${\mathbf Q}_\alpha$ that generate
the fibres. The latter satisfy the Lie bracket relation
\begin{eqnarray}
[{\mathbf Q}_\alpha, {\mathbf Q}_\beta] &=&-c^\gamma_{\;\;\alpha
\beta}
{\mathbf Q}_\gamma
\; ,
\label{(3)}
\end{eqnarray}
the $c$'s being the structure constants of the gauge group.

Gauge invariance of the metric (1) is the statement
\begin{eqnarray}
{\pounds_{\bf Q_{\alpha}}} \mbox{\boldmath $\gamma$} &=& 0
\; .
\label{(4)}
\end{eqnarray}
The ${\mathbf Q}_\alpha$ are Killing vector fields for the metric
$\mbox{\boldmath $\gamma$}$
and vertical vector fields on the space of
histories. The ${\mathbf Q}_\alpha$ and
$\mbox{\boldmath $\gamma$}$
together
define a unique connection $1$-form on this space:
\begin{eqnarray}
\mbox{{\boldmath $\omega$}}^{\alpha} &=& {\frak G}^{\alpha \beta} \bf
{Q}_{\beta} \cdot \mbox{\boldmath $\gamma$}
\; ,
\label{(5)}
\end{eqnarray}
where ${\frak G}^{\alpha \beta}$ is the Greens's function, appropriate
to the boundary conditions at hand (typically ``in-out'' or
``in-in''), of the operator
\begin{eqnarray}
{\frak F}_{\alpha \beta} &=& -\bf {Q}_{\alpha} \cdot \mbox{\boldmath
$\gamma$}
\cdot \bf{Q}_{\beta}
\; .
\label{(6)}
\end{eqnarray}
It is easy to see that $ {\mbox{\boldmath $\omega$}}^{\alpha} \cdot
 \bf{Q}_{\beta} = {\delta^{\alpha}}_{\beta}$ and that horizontal
 vectors are those that are perpendicular to the fibres. A horizontal
 vector may be obtained from any vector by applying  the {\it
 horizontal projection operator :}
\begin{eqnarray}
{\Pi^{i}}_{j} &=& {\delta^{i}}_{j} - {Q^{i}}_{\alpha}
{\omega^{\alpha}}_{j}
\; .
\label{(7)}
\end{eqnarray}

The trick for obtaining an effective action yielding ghost and
gauge-fixing-independent effective field equations is to make use of
the following connection on the space of
histories, first proposed by Vilkovisky \cite{2}:
\begin{eqnarray}
{\Gamma^{i}}_{jk}  &=&  {\Gamma_{\gamma}}^{i}{_{jk}}
- {Q^{i}}_{\alpha \cdot j} {\omega^{\alpha}}_{k}
- {Q^{i}}_{\alpha \cdot k} {\omega^{\alpha}}_{j} \nonumber \\
  &+& \frac{1}{2} {\omega^{\alpha}}_{j} {Q^{i}}_{\alpha \cdot l}
{Q^{l}}_{\beta} {\omega^{\beta}}_{k} + \frac{1}{2}
{\omega^{\alpha}}_{k} {Q^{i}}_{\alpha \cdot l}
{Q^{l}}_{\beta} {\omega^{\beta}}_{j}
\; .
\label{(8)}
\end{eqnarray}
$\Gamma_{\gamma}$ is the Riemannian connection associated with
$\mbox{\boldmath $\gamma$}$ and the dots denote covariant functional
differentiation based on it. Vilkovisky's connection has the following
properties:

\medskip

\noindent 1. {If a geodesic based on it is horizontal (vertical) at
one
point it is horizontal (vertical) throughout its length.}

\medskip

\noindent 2. {If $A$ is a gauge invariant functional
on the space of histories
then, for all $n$,
\begin{eqnarray}
A_{;(i_{1} \ldots i_{n})} &=& A._{(j_{1} \ldots j_{n})}
\Pi^{j_{1}}{_{i_{1}}} \ldots \Pi^{j_{n}}{_{i_{n}}}
\; ,
\label{(9)}
\end{eqnarray}
where the semicolon denotes covariant functional differentiation
based on
Vilkovisky's connection and parentheses indicate that a
symmetrization of the indices they embrace is to be performed.}
\medskip

Introduce a convenient base point $\varphi_*$ in the space of
histories,
e.g., Minkowski, Friedmann-Robertson-Walker, black hole. Let
$\varphi$ be an arbitrary point and $\lambda$ a Vilkovisky geodesic
connecting it to the base point, and let $s$ and $s_*$ be the values,
at $\varphi$ and $\varphi_*$ respectively, of an affine parameter
along $\lambda$. Define
\begin{eqnarray}
\mbox{\boldmath $\phi$} &=& (s -s_{*}) {\left (\frac{\partial}
{\partial s} \right )}_{\lambda (s_{*})}
\; .
\label{(10)}
\end{eqnarray}
$\mbox{\boldmath $\phi$}$ is a vector at $\varphi_{*}$, invariant
under rescaling of s. Its components $\phi^{a}$ in an arbitrary frame
at $\varphi_{*}$ may be called {\it geodesic normal fields}.

The $\phi^{a}$ can be used $[1,2]$ to carry out covariant functional
Taylor
expansions about $\varphi_{*}$, and in reference \cite{1} it is shown
that they
lead in a simple and straightforward way to the desired effective
action:
\begin{eqnarray}
\Gamma [\varphi_{*}, \bar{\phi}]
 &=&  S[\varphi_{*}, \bar{\phi}] + \Sigma [\varphi_{*}, \bar{\phi}]
\; .
\label{(11)}
\end{eqnarray}
Here $S$ is the classical action, $\bar \phi$ is the mean field, and
$\Sigma$ is the logarithm of a functional integral:
\begin{eqnarray}
&&\Sigma[\varphi_{*}, \bar{\phi}]=
\nonumber\\
&& - i {\left( {\rm ln} \int
e^{i(\frac{1}{2} F [\varphi_{*}, \bar{\phi}] \chi \chi + \frac{1}{6}
S_{3}[\varphi_{*}, \bar{\phi}] \chi \chi \chi + \ldots)}[d \chi]
\right)}_{\rm 1PI}
\; .
\label{(12)}
\end{eqnarray}
The subscript ``1PI'' means ``retain only $1$--particle--irreducible
graphs'', $S_{n}[\varphi_{*}, \bar{\phi}]$ denotes the $n$th
functional derivative of $S[\varphi_{*}, \bar{\phi}]$ with respect to
the $\bar \phi$'s, and $F$ has components given by
\begin{eqnarray}
F_{a b}[\varphi_{*},\bar{\phi}] & =&  S,_{a b}
[\varphi_{*},\bar{\phi}] +
\kappa_{\alpha \beta} P^\alpha_{\;\;a} P^\beta_{\;\;b}
\label{(13)}
\end{eqnarray}
where $(\kappa_{\alpha \beta})$ is any $\bar \phi$-independent
ultralocal invertible continuous matrix and $(P^\alpha_{\;\;a})$ is
any
$\bar \phi$-independent continuous matrix for which the operator
\begin{eqnarray}
{\hat {\frak F}}^\alpha_{\;\;\beta}&=& P^\alpha_{\;\;a}
 Q^a_{\;\;\beta}[\varphi_*, \bar \phi]
\label{(14)}
\end{eqnarray}
is nonsingular. The $Q^a_{\;\;\beta}$ are the components of ${\mathbf
Q}_\beta$ in the geodesic normal coordinate system.

The vertex functions  $S_{n}[\varphi_{*}, \bar{\phi}]$ have the
structure
\begin{eqnarray}
&&S,_{a_{1} \ldots a_{n}} [\varphi_{*}, \bar{\phi}]
 = \nonumber \\
&& \sum_{m = 0}^{\infty} \frac{1}{m!}
S_{;(a_{1} \ldots a_{n} b_{1} \ldots b_{m})} [\varphi_{*}]
\bar{\phi}^{b_{1}} \ldots \bar{\phi}^{b_{m}}
\; ,
\label{(15)}
\end{eqnarray}
and if one is computing amplitudes for physical processes taking
place in the background $\varphi_*$ these functions reduce to
\begin{eqnarray}
&&S,_{a_{1} \ldots a_{n}} [\varphi_{*}, 0] =
\nonumber \\
&& S._{{(b_{1} \ldots
b_{n})}} [\varphi_{*}] \Pi^{b_{1}}{_{a_{1}}} [\varphi_{*}] \ldots
\Pi^{b_{n}}{_{a_{n}}} [\varphi_{*}] \; ,
\label{(16)}
\end{eqnarray}
which are easily computed. The secret of ghost-free gauge theory is
simple: Replace all traditional vertex functions by (16)
and eliminate the ghost graphs. It is easy to show that every graph in
the loop-expansion of $\Sigma$ is individually invariant under both
gauge transformations and changes in the $P$'s and $\kappa$'s. This
result may seem surprising in view of the presence of the $P$'s and
$\kappa $'s in expression (13). However, do the following:
Proceed in the usual way, including the determinant of the ghost
operator (14) in the functional integral. Using dimensional
regularization discover that, because of the special forms taken by
the ghost vertices generated by the $Q^a_{\;\alpha}$, {\it every graph
containing a ghost line vanishes}!

The authors have applied this formalism to the Yang-Mills field, with
standard ``in-out'' boundary conditions. The renormalization program
proceeds as in the conventional approach, with these differences: The
$\phi^a$ are nonlocal composite operators and cannot serve directly as
interpolating fields for the S-matrix. To get from them to the local
fields $\varphi^i$ further renormalizations are required. The latter
do not spoil the asymptotic freedom of Yang-Mills theory but they lead
to slightly different $\beta$-functions for $\phi^a$ and $\varphi^i$.

If one is interested in expectation values rather than S-matrix
elements then the $\phi^a$ will serve as well as the $\varphi^i$. With
use of the ``closed-time-path'' formalism \cite{4}, which replaces the
$\phi^a$
by ``forward-time'' and ``backward-time'' fields $\phi^{\;\;a}_+$
and $\phi^{\;\;a}_-$
respectively, one can construct an ``in-in'' effective action that
generates the time evolution of expectation values $\langle \phi^a
\rangle$. The pole subtractions needed for the ``in-out'' effective
action suffice to renormalize the ``in-in'' effective action as
well. This is true for both gravity and Yang-Mills theory.

Although the computations for quantum gravity are algebraically far
more complicated than for Yang-Mills theory, Mirzabekian and
Vilkovisky \cite{5} have been able to show that $1$-loop contributions
to the
gravitational effective action already account for the Hawking
radiation from black holes, and they obtain general expression for the
relevant form factors describing back-reaction and decay. With a
ghost-and-gauge-fixing-independent effective action one can hope to
pin these form factors down so as to yield believable results in the
near Planckian regime. The ultimate aim is to solve the effective
field equations for $\langle \phi^a \rangle$, starting from
an imploding scalar-field shell and
following the dynamics through black hole formation and decay. Short
of a full blown string-theoretical analysis this is the only way of
resolving such issues as information loss or tunneling to other
universes.

\end{document}